# Calculation of coupling constant $g_{\phi\pi\gamma}$ in QCD sum rules

C. AYDIN [†] and A.H. YILMAZ [††]

*Physics Department, Karadeniz Technical University, 61080, Trabzon, Turkey*

## Abstract

The coupling constant of $\phi \rightarrow \pi^0 \gamma$ decay is calculated in the method of QCD sum rules. A comparison of our prediction on the coupling constant with the result obtained from analysis of the experimental data is performed.

---

[†] e-mail: coskun@ktu.edu.tr

[††] e-mail: hakany@ktu.edu.tr

The method of QCD sum rules [1] represent one of few methods capable of making predictions in the low- and medium-energy hadron physics starting basically from the QCD Lagrangian. The only phenomenological input parameters are the values of two or three quark and gluon condensates characterizing the properties of the physical vacuum. The field of application of the sum rules has been extended remarkably during the past twenty years [2]. Application of this method to polarization operators gives a determination of masses and couplings of low lying mesonic [1,3] and baryonic [4] states. The QCD sum rule method has been utilized to analyze many hadronic properties, and it yields an effective franework to investigate the hadronic observables such as decay constants and form factors within the nonperturbative contributions proprotional to the quark and gluon condensates. [2].

Radiative transitions between pseudoscalar (P) and vector (V) mesons have been an important subject in low-energy hadron physics for more than three decades. These transitions have been regarded as phenomenological quark models, potential models, bag models, and effective Lagrangian methods [5,6]. Among the characteristics of the strong interaction processes $g_{\phi\pi\gamma}$ coupling constant plays one of the most important roles, since they determine the strength of the hadron interactions. In the quark models, $V \to P + \gamma$ decays ($V = \phi, \rho, \omega$ ; $P = \pi, \eta, \eta'$) are reduced by the quark magnetic moment with transition $s = 1 \to s = 0$, where $s$ is the total spin of the $q\bar{q}$ − system (in the corresponding meson). Actually these quantities can be calculated directly from QCD. Low-energy hadron interactions are governed by nonperturbative QCD so that it is very difficult to get the numerical values of the coupling constants from the first principles. For that reason a semiphenomenological method of QCD sum rules can be used, which nowdays is the standart tool for studying of various characteristics of hadron interactions. On the other hand, vector meson-pseudoscalar meson-photon $VP\gamma$-vertex also plays a role in photoproduction reactions of vector mesons on nucleons. It should be notable that in these decays ($V \to P\gamma$) the four-momentum of the pseudoscalar meson $P$ is time-like, $P'^2 > 0$, whereas in the pseudoscalar exchange amplitude contributing to the photoproduction of vector mesons it is space-like $P'^2 < 0$. Therefore, it is of interest to study the effective coupling constant $g_{VP\gamma}$ from another point of view as well. The coupling constants of $\rho^0 \to \pi_0\gamma$ and $\omega \to \pi^0\gamma$ decays were calculated by Gökalp and Yilmaz [5] in the QCD sum rules. In this paper we apply the QCD sum rules to determine the $g_{\phi\pi\gamma}$ coupling constant.

In the following, we will analyse radiative $\phi \to \pi^0 \gamma$ decay using QCD sum rules. We begin with the observation that the $\phi \to \pi^0 \gamma$ decay width vanishes if the $\phi$ meson in pure $s\bar{s}$ state. The measured width $\Gamma(\phi \to \pi^0 \gamma) = (5.4 \pm 0.16)$ keV [7] is significantly different from zero which is primarily due to $\omega\phi-$ mixing. Bramon et al. [8] have recently studied radiative $VP\gamma$ transition between vector ($V$) mesons and pseudoscalar ($P$) mesons, and using the available experimental information they have determined the mixing angle as well as other relevant parameters for the $\omega - \phi$ system and also for the physical $\omega$ and $\phi$ meson states as

$$|\omega> = \cos\beta |\omega_0> - \sin\beta |\phi_0>$$
$$|\phi> = \sin\beta |\omega_0> + \cos\beta |\phi_0>$$

where $|\omega_0> = \frac{1}{\sqrt{2}} |u\bar{u} + d\bar{d}>$ and $|\phi_0> = |s\bar{s}>$ are the non-strange and the strange basis states. The mixing angle that used is determined by Bramon et al. as $\beta = (3.4 \pm 0.2)°$ [8]. We, therefore, choose the interpolating currents for $\omega$ and $\phi$ mesons defined in the quark flavors basis as

$$j_\mu^\omega = \cos\beta j_\mu^{\omega_0} - \sin\beta j_\mu^{\phi_0}$$
$$j_\mu^\phi = \sin\beta j_\mu^{\omega_0} + \cos\beta j_\mu^{\phi_0}$$

where $j_\mu^{\omega_0} = \frac{1}{6}(\bar{u}\gamma_\mu u + \bar{d}\gamma_\mu d)$ and $j_\mu^{\phi_0} = -\frac{1}{3}\bar{s}\gamma_\mu s$.

According to the general strategy of QCD sum rules method, the coupling constants can be calculated by equating the representations of a suitable correlator calculated in terms of hadronic and quark-gluon degrees of freedom. In order to do this we consider the following correlation function by using the appropriately chosen currents

$$\Pi_{\mu\nu}(p,p') = \int d^4x\, d^4y\, e^{ip'.y} e^{-ip.x} <0|T\{J_\mu^\gamma(0) J_\nu^\phi(x) J_s(y)\}|0> . \tag{1}$$

We choose the interpolating current for the $\phi$ and $\pi$ mesons as

$$J_\mu^\phi = \sin\beta j_\mu^{\omega_0} = \frac{1}{6}(\bar{u}\gamma_\mu u + \bar{d}\gamma_\mu d)\sin\beta, \quad \text{and} \quad J_\mu^{\pi^0} = \frac{1}{2}[\bar{u}(i\gamma_5)u - \bar{d}(i\gamma_5)d] \quad \text{respectively.}$$

$J_\mu^\gamma = e_u(\bar{u}\gamma_\mu u) + e_d(\bar{d}\gamma_\mu d)$ is the electromagnetic current with $e_u$ and $e_d$ being the quark charges.

The theoretical part of the sum rule in terms of the quark-gluon degrees of freedom for the coupling constant $g_{V\pi\gamma}$ is calculated by considering the perturbative contribution and the power corrections from operators of different dimensions to the three-point correlation function $\Pi_{\mu\nu}$. For the perturbative contribution we study the lowest order bare-loop diagram. Moreover, the power corrections from the operators of different dimensions $<\bar{q}q>$, $<\bar{q}\eta.Gq>$, and $<(\bar{q}q)^2>$ are considered in the work. Since it is estimated to be negligible for light quark systems, we did not consider the gluon condensate contribution proportional to $<G^2>$. We performe the calculations of the power corrections in the fixed point gauge [9]. We also work in the limit $m_q = 0$ and in this limit the perturbative bare-loop diagram does not make any contribution. In fact, by considering this limit only operators of dimensions d=3 and d=5 make contributions which are proportional to $<\bar{q}q>$ and $<\bar{q}\eta.Gq>$, respectively. The relevant Feynman diagrams for power corrections are given in Fig. 1.

On the other hand, in order to calculate the phenomenological part of the sum rule in terms of hadronic degrees of freedom, a double dispersion relation satisfied by the vertex function $\Pi_{\mu\nu}$ is considered [1-3]:

$$\Pi_{\mu\nu}(p,p') = \frac{1}{\pi^2}\int ds_1 \int ds_2 \frac{\eta_{\mu\nu}(s_1,s_2)}{(p^2-s_1)(p'^2-s_2)} \quad (2)$$

where we ignore possible substruction terms since they will not make any contributions after Borel transformation. For our purpose we choose the vector and pseudoscalar channels and saturating this dispersion relation by the lowest lying meson states in these channels the physical part of the sum rule is obtained as

$$\Pi_{\mu\nu}(p,p') = \frac{<0|J_\nu^\phi|\phi><\phi(p)|J_\mu^\gamma|\pi(p')><\pi|J_s|0>}{(p^2-m_\phi^2)(p'^2-m_\pi^2)} + ..., \quad (3)$$

where the contributions from the higher states and the continuum are given by dots. The overlap amplitudes for vector and pseudscalar mesons are $<0|J_\mu^\phi|\phi>=\lambda_\phi\varepsilon_\mu^\phi$, where $\varepsilon_\mu^\phi$ is the polarization vector of the vector meson and $<\pi|J_s|0>=\lambda_s$, respectively. The matrix element of the electromagnetic current is given by

$$<\phi(p)|J_\mu^\gamma|\pi(p')>=-i\frac{e}{m_\phi}g_{\phi\pi\gamma}K(q^2)\varepsilon^{\mu\alpha\beta\delta}P_\alpha u_\beta q_\delta \quad (4)$$

where $q = p - p'$ and $K(q^2)$ is a form factor with $K(0)=1$. This matrix element defines the coupling constant $g_{\phi\pi\gamma}$ by means of the effective Lagrangian

$$\mathcal{L}=\frac{e}{m_\phi}g_{\phi\pi\gamma}\varepsilon^{\mu\nu\alpha\beta}(\partial_\mu\phi_\nu\partial_\nu A_\beta\pi^0) \quad (5)$$

describing the $\phi\pi\gamma$ – vertex [10].

After performing the double Borel transform with respect to the variables $Q^2 = -p^2$ and $Q'^2 = -p'^2$, and by considering the gauge-invariant structure $\varepsilon^{\mu\alpha\beta\delta} p_\alpha q_\delta$, we obtain the sum rule for the coupling constant as

$$g_{\phi\pi\gamma} = \frac{m_\phi}{\lambda_\phi \lambda_\pi} e^{m_\phi^2/M^2} e^{m_\pi^2/M'^2} \left( e_u < \bar{u}u > - e_d < \bar{d}d > \right) \left( -\frac{3}{4} + \frac{5}{32} \frac{m_0^2}{M^2} - \frac{3}{32} \frac{m_0^2}{M'^2} \right) \sin\beta \quad (6)$$

where we used the relation $< \bar{q}\eta.Gq >= m_0^2 < \bar{q}q >$. In our calculations we use the numerical values of $m_0^2 = (0.8 \pm 0.02)\,\text{GeV}^2$, $<\bar{u}u>=<\bar{d}d>=(-0.014 \pm 0.002)\,\text{GeV}^3$, $m_\phi = 1.02\,\text{GeV}$, $m_{\pi^0} = 0.138\,\text{GeV}$ [2]. Since $\lambda_\pi = f_\pi \frac{m_\pi^2}{m_u + m_d}$, then $\lambda_\pi$ is given as $\lambda_\pi = 0.18\,\text{GeV}^2$. We note that neglecting the electron mass the $e^+e^-$ decay width of $\phi$ meson is given as $\Gamma(\phi \to e^+e^-) = \frac{4\pi\alpha^2}{3} \frac{\lambda_\phi^2}{m_\phi^3}$. Then using the value from the experimental leptonic decay width of $\phi$ [4], we obtain the value $\lambda_\phi = (0.079 \pm 0.001)\,\text{GeV}^2$ for the overlap amplitude $\phi$ meson. In order to examine the dependence of $g_{\phi\pi\gamma}$ on the Borel masses $M^2$ and $M'^2$, we chose $M'^2 = 1$, 1.2 and $1.4\,\text{GeV}^2$. Since the Borel mass $M^2$ is an auxiliary parameter and the physical quantities should not depend on it, one must look for the region where $g_{\phi\pi\gamma}$ is practically independent of $M^2$. We determined that this condition is satisfied in the interval $1\,GeV^2 \leq M^2 \leq 1.6\,GeV^2$. The variation of the coupling constant $g_{\phi\pi\gamma}$ as a function of Borel parameters $M^2$ for different values of $M'^2$ are shown in Figs. 2 and 3. Examination of this figure point out that the sum rule is rather stable with these reasonable variations of $M^2$ and $M'^2$. We choose the middle value $M^2 = 1\,\text{GeV}^2$ for the Borel parameter in its interval of variation and obtain the coupling constant $g_{\phi\pi\gamma}$ as $g_{\phi\pi\gamma} = 0.125 \pm 0.002$, where only the error arising from the numerical analysis of the sum rule is considered. If we take the uncertainties into account by a conservative estimation, we have the coupling constant $g_{\phi\pi\gamma} = 0.125 \pm 0.004$.

We would like to compare our predictions on $g_{\phi\pi\gamma}$ with experimental result. The decay width of $\phi \to \pi^0\gamma$ decay is given by

$$\Gamma(\phi \to \pi^0 \gamma) = \frac{\alpha}{24} \frac{(m_\phi^2 - m_\pi^2)^3}{m_\phi^5} g_{\phi\pi\gamma}^2. \tag{7}$$

Using the experimental value of $\Gamma(\phi \to \pi^0 \gamma)$ [7], the $g_{\phi\pi\gamma}$ coupling constant is obtained from above equation as $g_{\phi\pi\gamma} = 0.135 \pm 0.002$ which is very close to the sum rule estimation.

The result we have obtained by the QCD sum rule utilizing three point correlation functions, agrees with the values obtained from an analysis of experimental decay width and therefore supplements the study of this decay using phenemenological considerations.

ACKNOWLEDGMENT
We like to thank Profs. A. Gökalp and O. Yilmaz for valuable discussions and comments.

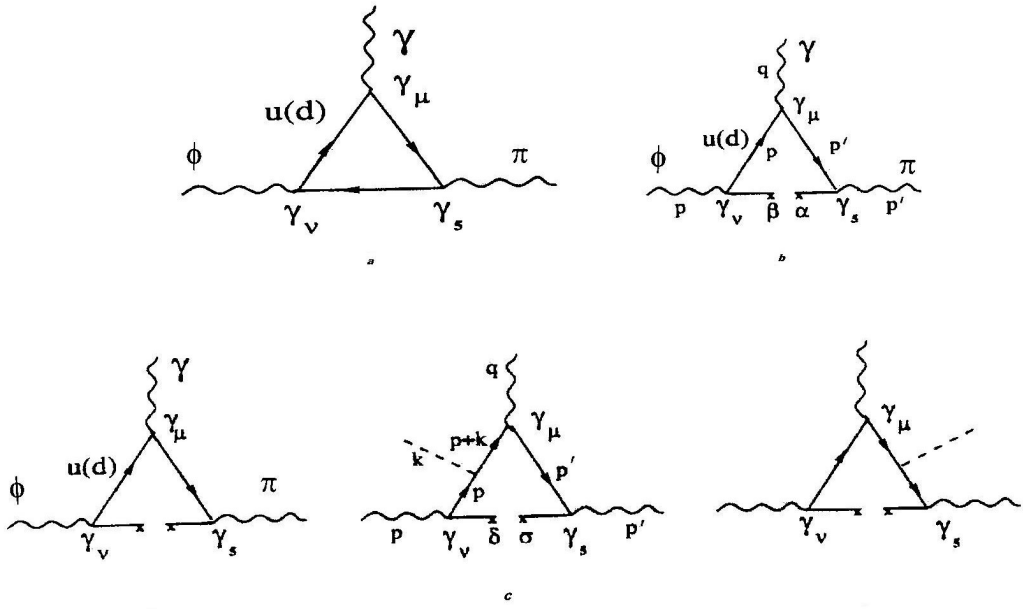

Figure 1. a-c) Feynman diagrams for the $\phi\pi\gamma$-vertex: a) bare loop diagram, b) d=3 operator corrections, and c) d=5 operator corrections. The dotted lines denote gluons.

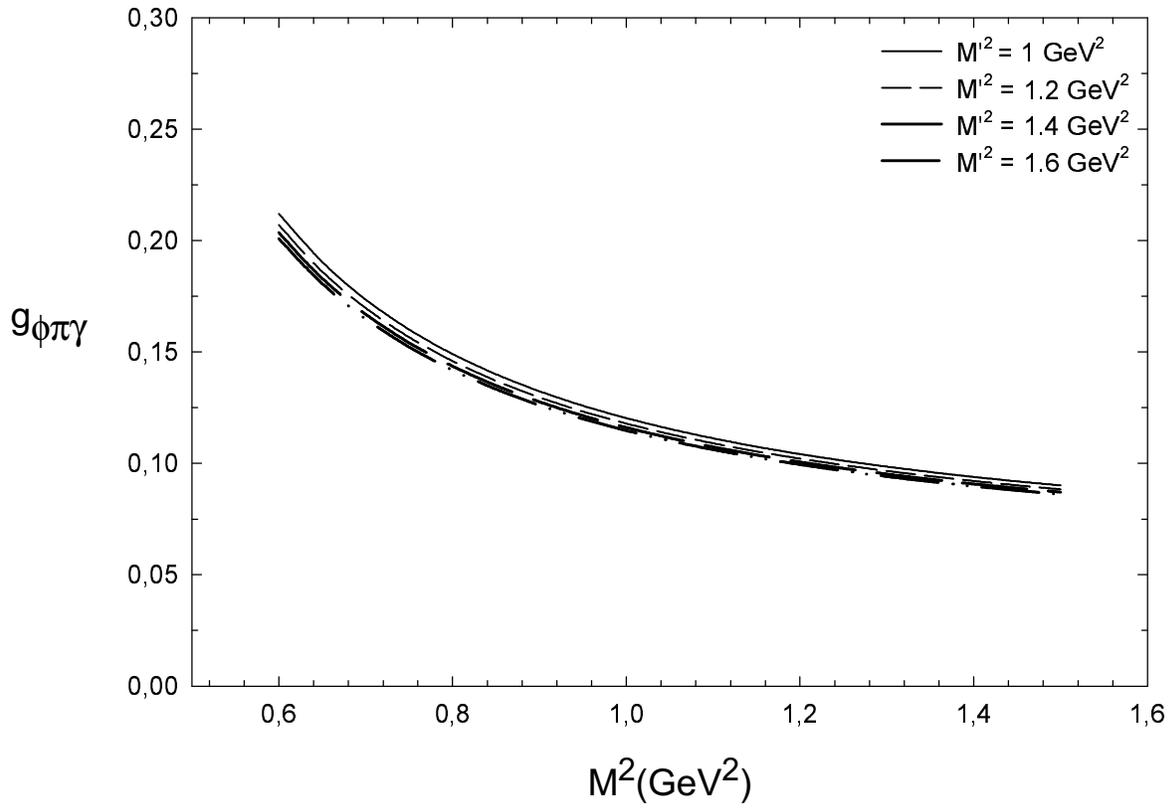

Figure 2. The coupling constant $g_{\phi\pi\gamma}$ as a function of the Borel parameter $M^2$ for different values of $M'^2$.

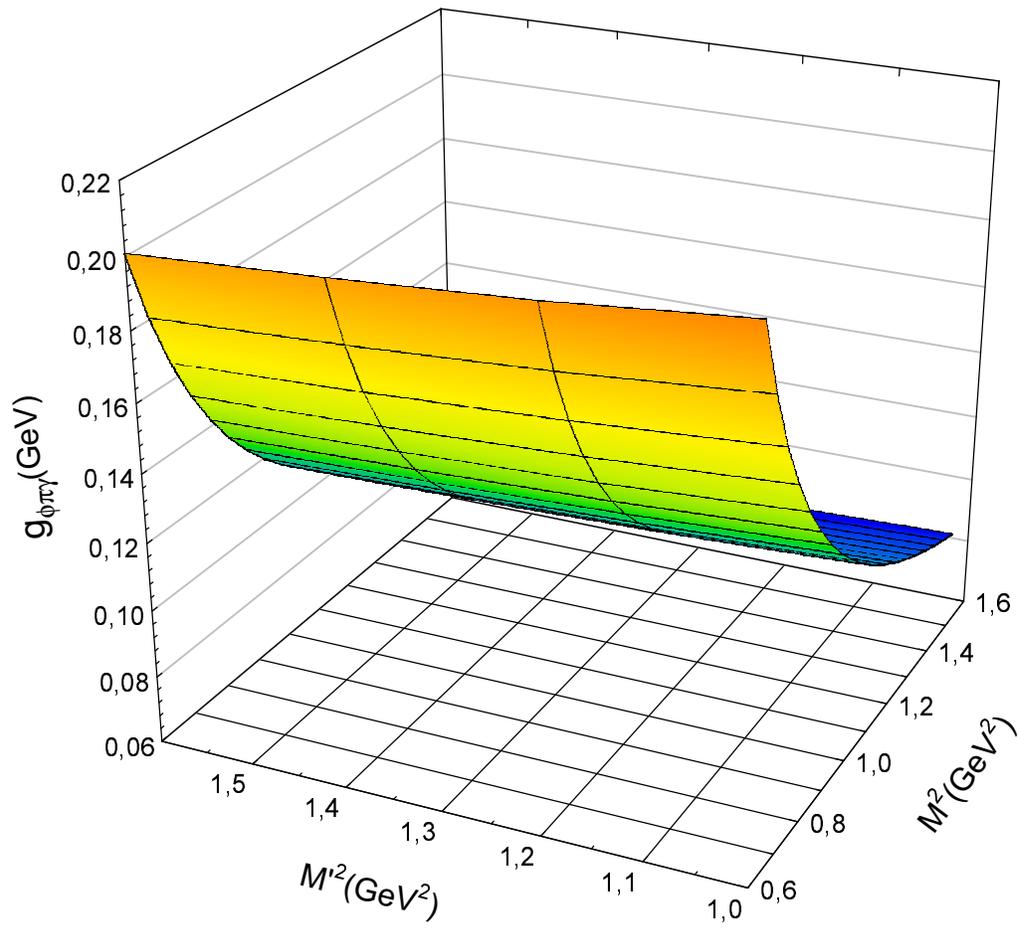

Figure 3. Coupling $g_{\phi\pi\gamma}$ as a function of the Borel parameters $M^2$ for different values of $M'^2$.